\documentclass{mn2e}
\usepackage[dvips]{color}
\usepackage{epsf}

\title[V838 Mon: light echo evolution and distance estimate]
{V838 Mon: light echo evolution and distance estimate}

\author[L. A. Crause et al.]
{Lisa A. Crause,$^{1,2}$\thanks{E-mail: lisa@saao.ac.za; w.lawson@adfa.edu.au; 
jwm@saao.ac.za; fm@saao.ac.za} Warrick A. Lawson,$^{3\star}$ John W. 
Menzies$^{1\star}$ and Fred Marang$^{1\star}$\\
$^{1}$South African Astronomical Observatory, P.O. Box 9, Observatory 7935, 
South Africa \\
$^{2}$Department of Astronomy, University of Cape Town, Private Bag,
Rondebosch 7700, South Africa\\
$^{3}$School of Physical, Environmental and Mathematical Sciences, University 
of New South Wales, Australian Defence Force Academy, \\
Canberra, ACT 2600, Australia \\}

\date{Accepted 2005 ...... Received 2005 ...... in original form 2004 ......}

\pagerange{\pageref{firstpage}--\pageref{lastpage}}
\pubyear{2004}

\begin{document}

\maketitle

\label{firstpage}

\begin{abstract}

Following its 2002 February eruption, V838 Mon developed a light echo that
continues to expand and evolve as light from the outburst scatters off
progressively more distant circumstellar and/or interstellar material. 
Multi-filter images of the light echo, obtained with the South African
Astronomical Observatory (SAAO) 1.0-m telescope between 2002 May and 2004
December, are analysed and made available electronically.  The expansion of
the light echo is measured from the images and the data compared with models
for scattering by a thin sheet and a thin shell of dust.  From these model
results we infer that the dust is likely in the form of a thin sheet distant
from the star, suggesting that the material is of interstellar origin, rather 
than being from earlier stages in the star's evolution.  Although the fit is 
uncertain, we derive a stellar distance of $\sim 9$ kpc and a star-dust 
distance of $\sim 5$ pc, in good agreement with recent results reported from 
other methods.  We also present {\it JHKL\,} and Cousins {\it UBVRI\,} 
photometry obtained at the SAAO during the star's second, third and fourth 
observing seasons post-outburst.  These data show complex infrared colour 
behaviour while V838 Mon is slowly brightening in the optical.

\end{abstract}

\begin{keywords}
stars: individual: V838 Mon --- stars: photometry --- stars: circumstellar 
matter --- stars: interstellar matter --- stars: distances --- stars: imaging
\end{keywords}

\section{Introduction}

V838 Mon appeared on the scene with a nova-like outburst of amplitude
$\Delta V \approx 6$ magnitudes in 2002 January (Brown 2002) and its
extraordinary subsequent evolution indicates that this object represents an
entirely new type of variable (Munari et al. 2002, Evans et al. 2003). 
Besides its unique light curve, V838 Mon also underwent dramatic spectral
evolution as the star rapidly expanded and cooled, transforming itself from
an object vaguely resembling a mid-K giant into what Evans et al. (2003)
propose to be the first known L supergiant.  The emergence of a blue
continuum in spectra obtained in 2002 October (Desidera \& Munari 2002,
Wagner \& Starrfield 2002) suggests that V838 Mon may be a binary, although
this early-type star could possibly be a line-of-sight companion.  Various
unconventional models, including merging main sequence stars (Soker \&
Tylenda 2004) and an expanding red giant engulfing planets (Retter \& Marom
2003), have been proposed to explain the star's behaviour.  More recently,
van Loon et al. (2004) argued that V838 Mon may be a single, low-mass AGB
star experiencing thermal pulses while Munari et al. (private communication)
and Tylenda, Soker \& Szczerba (2004) both favour a young binary system with
a massive progenitor for the outbursting component.  These diverse
interpretations of the 2002 phenomena indicate the depth of the mystery
surrounding this unusual variable.

Optical imaging 2 weeks after V838 Mon's 2002 February eruption revealed the
development of a light echo (Henden, Munari \& Schwartz 2002); the result of
light from the outburst being scattered into our line of sight by previously
unseen circumstellar and/or interstellar material.  We see the echo
expanding as the light reaches progressively more distant dust.  The only
other known Galactic light echo was associated with Nova Persei in 1901
(Couderc 1939) and hence the V838 Mon echo has received widespread
attention, particularly following the release of spectacular {\it Hubble
Space Telescope} ({\it HST}) images\footnote{See {\tt
http://hubblesite.org/newscenter/newsdesk/archive/
releases/2004/10/image/}}.

Aesthetics aside, the light echo also provides important information about
this enigmatic star.  The observed expansion rate of the echo will
eventually yield the nature of the scattering material, as well as the
distance to the star and the distance between the star and the dust.
Knowledge of V838 Mon's luminosity will allow stricter evaluation of the
various models that attempt to account for the star's 2002 eruptions. 
Furthermore, since the light echo illuminates progressively more distant
material, each image provides a once-off map of the dust distribution on a
given light echo paraboloid.  Assembling these maps, taking into account the
transformations required to correct for projection effects, would ultimately
provide a 3-dimensional reconstruction of the scattering dust.

\section{Observations}

V838 Mon was included in the SAAO Infrared Service Programme following the
initial outburst in 2002 January.  Data from the first observing season were
presented by Munari et al. (2002) and Crause et al. (2003).  Subsequent {\it
JHKL\,} measurements, also made with the IRP Mark II photometer on the
0.75-m telescope, are listed in Table 1.  As for the first season data, a 36
arcsec aperture was used.  Target observations were interspersed with Carter
infrared standards (Carter 1995) situated at similar airmasses and the data
reduced with SAAO software that interpolates between standard stars to
calibrate the sky-subtracted measurements.  Given the infrared brightness of
V838 Mon, contributions from faint field stars within the aperture are
negligible.  The typical uncertainty in the corrected magnitudes is 0.03 mag 
for the {\it JHK\,} data and 0.05 mag for the $L$ band observations.

All of the optical images and photometry presented here were obtained with
the SAAO 1.0-m telescope and 1k $\times$ 1k STE4 CCD camera.  The CCD was
used without prebinning, resulting in a plate scale of 0.31 arcsec
pixel$^{-1}$.  During 12 observing runs between 2002 May and 2004 December,
imaging was limited to moonless conditions with seeing typically better than
2 arcsec.  The CCD frames were bias-subtracted, flat-fielded and trimmed
using the {\tt ccdproc} package within {\tt IRAF}.

Whenever possible, V838 Mon observations were bracketed with red [$(V-I)$ 
$\sim 2$] Landolt standard stars, but the extreme post-maximum colours [($V-R$) 
$\sim 3$ and ($V-I$) $\sim 6$] required large extrapolations for the
standard transformations.  PSF-modeling was performed with SAAO software and
the resulting Cousins {\it UBVRI\,} photometry is listed in Table 2.  The
faintness of the star and the low sensitivity of the CCD in the $U$ band
made it impractical to obtain high signal-to-noise data and so $U$
magnitudes are only quoted to one decimal place.  Having obtained calibrated
photometry at least once during most runs, it was possible to differentially
correct additional observations made under non-photometric conditions using
field stars located near V838 Mon.  These differential measurements are
marked with stars in Table 2.  Due to V838 Mon's extreme redness, the
short exposures required to prevent saturation limited the accuracy of the
$I$ band photometry as all comparison stars were severely underexposed.

Throughout this paper we use Julian Dates in abbreviated form, JD--2452000
(AJD).  In Section 6 we introduce day numbers that indicate the interval
between the 2002 February 5 (AJD 311) outburst peak and the observation in
question.

\section{Photometry}

\begin{table}
\begin{center}
\caption{SAAO {\it JHKL\,} photometry of V838 Mon obtained with the 0.75-m 
telescope between 2002 September 8 and 2004 December 10.  Dates are given as 
JD--2452000 (AJD) and horizontal bars separate the observations made during 
the second, third and early-fourth observing seasons.}
\begin{tabular}{ccccc}
\hline
AJD & $J$ & $H$ & $K$ & $L$ \\
\hline
525.663 & 6.72 & 5.35 & 4.32 & 3.05 \\
529.666 & 6.75 & 5.35 & 4.32 & 3.08 \\
534.646 & 6.72 & 5.39 & 4.34 & 3.03 \\
555.555 & 6.64 & 5.32 & 4.30 & 2.96 \\
563.636 & 6.56 & 5.29 & 4.30 & 2.93 \\
567.631 & 6.56 & 5.30 & 4.30 & 2.93 \\
568.619 & 6.54 & 5.31 & 4.30 & 2.94 \\
572.627 & 6.54 & 5.30 & 4.30 & 2.93 \\
574.632 & 6.54 & 5.30 & 4.31 & 2.93 \\
593.555 & 6.51 & 5.31 & 4.32 & 2.93 \\
603.597 & 6.52 & 5.32 & 4.34 & 2.92 \\
661.413 & 6.73 & 5.53 & 4.59 & 3.08 \\
667.421 & 6.71 & 5.54 & 4.62 & 3.11 \\
688.377 & 6.80 & 5.65 & 4.78 & 3.26 \\
696.403 & 6.76 & 5.65 & 4.79 & 3.26 \\
702.371 & 6.82 & 5.70 & 4.86 & 3.35 \\
753.208 & 6.84 & 5.70 & 4.92 & 3.50 \\
761.227 & 6.84 & 5.71 & 4.96 & 3.51 \\
775.201 & 6.86 & 5.72 & 4.99 & 3.55 \\
\hline
894.672 & 6.95 & 5.78 & 5.09 & 3.62 \\
933.627 & 7.00 & 5.82 & 5.13 & 3.68 \\
955.596 & 7.03 & 5.86 & 5.16 & 3.68 \\
969.598 & 7.05 & 5.86 & 5.17 & 3.73 \\
974.588 & 7.08 & 5.87 & 5.17 & 3.73 \\
1031.443 & 7.10 & 5.88 & 5.22 & 3.80 \\
1063.323 & 7.08 & 5.89 & 5.25 & 3.83 \\
1070.316 & 7.13 & 5.89 & 5.24 & 3.80 \\
1139.207 & 7.20 & 5.92 & 5.27 & 3.89 \\
1142.198 & 7.20 & 5.93 & 5.27 & 3.86 \\
\hline
1277.641 & 7.47 & 6.15 & 5.45 & 4.13 \\
1290.635 & 7.43 & 6.12 & 5.43 & 4.10 \\
1327.591 & 7.41 & 6.12 & 5.44 & 4.10 \\
1350.534 & 7.49 & 6.18 & 5.47 & 4.03 \\
\hline
\end{tabular}
\end{center}
\end{table}

\begin{table}
\caption{SAAO {\it UBVRI\,} photometry of V838 Mon obtained between 2002 
September 12 and 2004 December 12.  Low signal-to-noise data are only quoted 
to 1 decimal place.  Dates marked with a star indicate differentially 
corrected data from non-photometric nights and horizontal bars separate the
observations made during the second, third and early-fourth observing seasons.}
\begin{center}
\begin{tabular}{lccccc}
\hline
AJD & $V$ & ($B-V$) & ($U-B$) & ($V-R$) & ($V-I$) \\
\hline
529.63 & 16.28 & 0.4 & -- & 2.00 & 5.71 \\
$530.61^{\star}$ & 16.30 & 0.4 & -- & 2.02 & -- \\
$531.61^{\star}$ & 16.29 & 0.4 & -- & 2.03 & 5.72 \\
$533.63^{\star}$ & 16.28 & -- & -- & 2.02 & 5.71 \\
$534.62^{\star}$ & 16.27 & -- & -- & 2.00 & 5.70 \\
$577.56^{\star}$ & 16.24 & 0.6 & -- & 2.76 & 6.43 \\
578.51 & 16.25 & 0.6 & --0.1 & 2.74 & 6.44 \\
$579.56^{\star}$ & 16.26 & 0.5 & -- & 2.74 & 6.40 \\
$580.57^{\star}$ & 16.25 & 0.5 & -- & 2.72 & -- \\
581.54 & 16.27 & 0.5 & -- & 2.80 & 6.47 \\
661.38 & 16.27 & 0.49 & $-0.1$ & 2.85 & 6.42 \\
662.31 & 16.24 & 0.51 & $-0.1$ & 2.86 & 6.47 \\
663.33 & 16.25 & 0.50 & $-0.1$ & 2.86 & 6.46 \\
664.34 & 16.25 & 0.51 & $-0.1$ & 2.86 & 6.44 \\
667.32 & 16.25 & 0.53 & $-0.1$ & 2.87 & 6.41 \\
745.27 & 16.15 & 0.57 & 0.0 & 3.03 & 6.41 \\
$749.25^{\star}$ & 16.12 & 0.56 & -- & -- & 6.31 \\
$751.27^{\star}$ & 16.14 & 0.54 & -- & 3.06 & 6.36 \\
$780.22^{\star}$ & 16.13 & 0.63 & -- & 3.17 & -- \\
781.22 & 16.13 & 0.63 & -- & 3.03 & 6.33 \\
$783.21^{\star}$ & 16.13 & 0.62 & -- & -- & -- \\
$784.21^{\star}$ & 16.13 & 0.62 & 0.0 & 3.06 & 6.35 \\
\hline
$893.62^{\star}$ & 15.65 & -- & -- & 3.08 & 6.08 \\
894.64 & 15.65 & -- & -- & 3.07 & 6.12 \\
$898.61^{\star}$ & 15.65 & 0.88 & -- & 3.08 & 6.05 \\
$947.50^{\star}$ & 15.64 & -- & -- & 2.94 &  -- \\
$1133.25^{\star}$ & 15.47 & -- & -- & 2.67 & 5.85 \\
1134.23 & 15.47 & -- & -- & 2.65 & 5.86 \\
$1135.26^{\star}$ & 15.42 & 1.12 & -- & 2.62 & 5.81 \\
\hline
$1347.46^{\star}$ & 15.18 & -- & -- & -- & -- \\
$1348.42^{\star}$ & 15.18 & 0.73 & -- & 2.28 & 4.71 \\
$1349.50^{\star}$ & 15.18 & -- & -- & -- & 4.71 \\
1350.44 & 15.18 & 0.74 & 0.7 & 2.29 & 4.71 \\
$1351.46^{\star}$ & 15.19 & -- & -- & -- & -- \\
$1352.47^{\star}$ & 15.19 & 0.73 & -- & -- & 4.72 \\
\hline
\end{tabular}
\end{center}
\end{table}

\begin{figure}
\epsfxsize=8.0cm
\epsffile{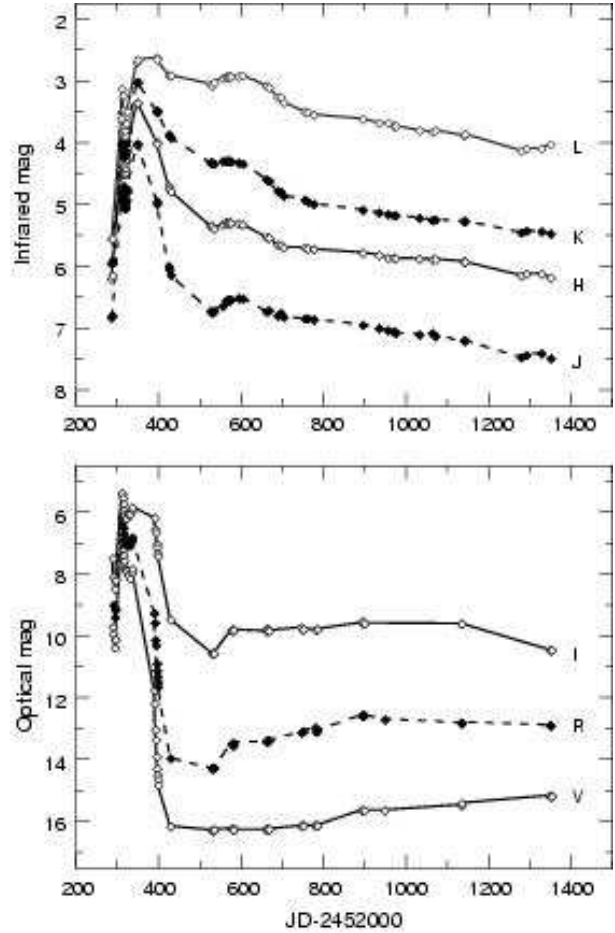}
\caption{Infrared ($JHKL$) and optical ($VRI$) light curves for V838 Mon, 
including first season data from Crause et al. (2003) and second, third and
early-fourth season SAAO data reported in Tables 1 and 2.  The magnitude scale 
for the infrared (upper) panel is twice that for the optical (lower) data.  
Note the bump at $\sim 600$ d visible in all bands except $V$, as well as the 
gradual brightening in the $V$ band since AJD $\sim 800$.}
\end{figure}

V838 Mon's complete $VRI$ and $JHKL$ light curves are shown in Fig. 1.  The
$U$ and $B$ data are relatively sparse and hence were not included in this
figure.  Note that, to preserve detail in the $JHKL$ curves, the vertical
scale of the infrared (top) panel is twice that of the optical (bottom)
panel.  Arbitrary alternating symbols were used to help distinguish the
various bands.

During the 4 month interval between the peak of the 2002 outbursts and the
star entering conjunction, V838 Mon faded by $\sim$ 9, 8 and 4 magnitudes in
$V$, $R$ and $I$ respectively.  The infrared brightness peaked $\sim 40$
days after the optical maximum and the trend of the fade amplitude
decreasing with wavelength continued through the $JHKL$ bands.  Crause et
al. (2003) interpreted this first season photometric behaviour as being
indicative of a brief dust formation episode following the third optical
outburst, however, this has not been confirmed spectroscopically.

Although the star's brightness was decreasing in all bands in 2002 June (AJD
$\sim 430$), V838 Mon reappeared in 2002 September (AJD $\sim 530$) with all
but the $V$ band showing a renewed increase.  The $R$ and $I$ bands both
brightened by $\sim 0.8$ magnitudes before approximately levelling off at
13.5 and 9.8 respectively while the $JHKL$ light curves developed a distinct
broad peak roughly centred on AJD 600.  The onset of this peak corresponds
to the time when Evans et al. (2003) classified V838 Mon's spectrum as that
of an L supergiant and the variation in the shape of this bump across the 4
infrared bands, becomming broader and less peaked from $J$ through $L$,
reflects the dominance of the cool component.

Since the star remained near its minimum optical brightness throughout the
second season, the constant photometric contribution from the blue companion
can be seen in the ($U-B$) column of Table 2.  The steady brightening in the
$V$ band during the third season is a continuation of the behaviour reported
by Crause (2003) when the star emerged from conjunction in 2003 September. 
A consequence of this brightening is that the ($V-R$) and ($V-I$) colours
became bluer while the ($B-V$) colour continued to redden.

\begin{figure*}
\epsfxsize=16.0cm
\epsffile{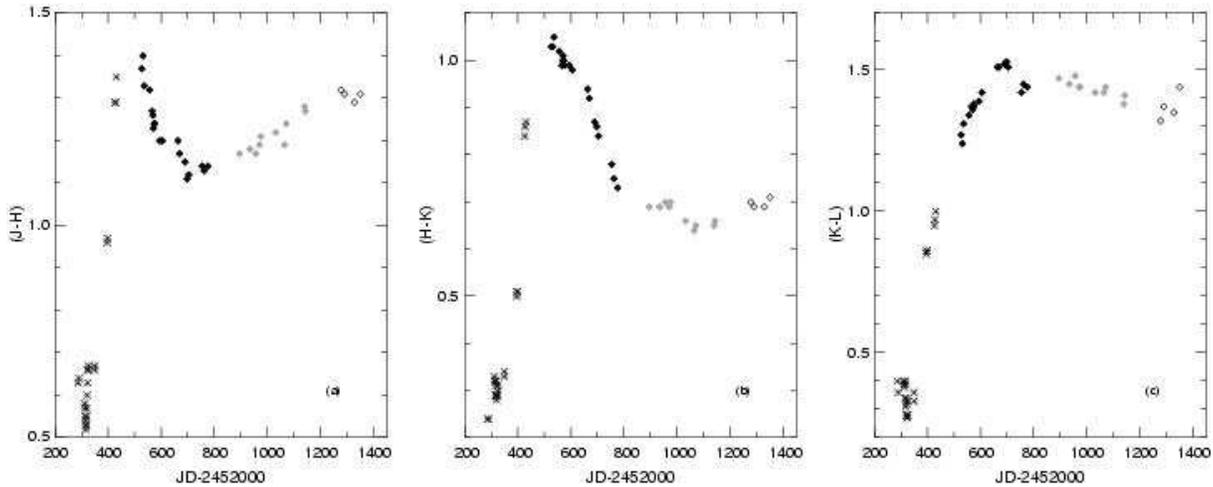}
\caption{($J-H$), ($H-K$) and ($K-L$) colour curves for V838 Mon.  Asterisks
indicate published data from the first observing season while black, grey and 
open diamonds mark new data from the second, third and early-fourth seasons 
respectively.}
\end{figure*}

\begin{figure*}
\epsfxsize=16.0cm
\epsffile{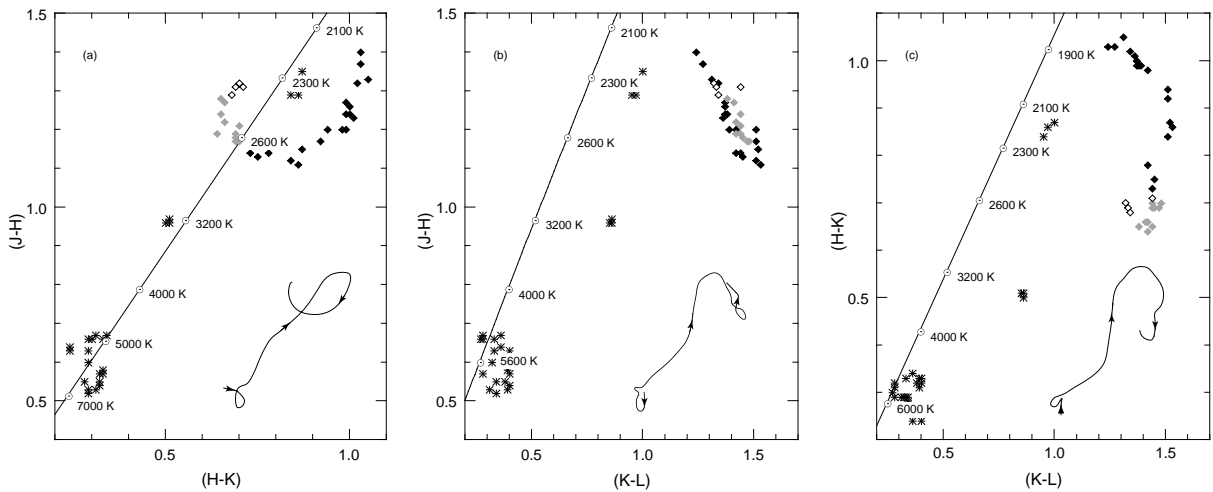}
\caption{($J-H$) vs ($H-K$), ($J-H$) vs ($K-L$) and ($H-K$) vs ($K-L$)
colour-colour diagrams showing the complex infrared evolution of V838 Mon. 
Symbols as in Fig. 2.  The locus of blackbody colours reddened by $A_{V} =
2.8$ mag, as well as curves illustrating the temporal evolution of the data, 
are included in each panel.  See Crause et al. (2003) for details of the 2002 
infrared behaviour.}
\end{figure*}

The infrared ($J-H$), ($H-K$) and ($K-L$) colour curves for V838 Mon are
shown in Fig. 2 and the various infrared colour-colour diagrams in Fig. 3. 
Asterisks mark data from the first season (reported by Munari et al. 2002 
and Crause et al. 2003) while diamond symbols indicate those presented in 
this paper.  Data from the second, third and early-fourth seasons are shown 
as black, grey and open symbols respectively.  The gaps roughly centred on 
AJD 460, 830 and 1200 (2002, 2003 and 2004 July) are due to the star being 
in conjunction with the Sun.

Figs 2a and 2b show the star recovering from its reddest ($J-H$) and ($H-K$)
colours during the second season (2002 September to 2003 May; AJD $526-776$)
while Fig. 2c reveals that the ($K-L$) colour only reached its reddest value
about 170 d after the ($J-H$) and ($H-K$) colours did.  Since the third 
season (2003 September; AJD 895), the ($J-H$) curve showed a redward trend
while the ($K-L$) colour became slightly bluer.  

V838 Mon's complex colour behaviour is further illustrated in Fig. 3. 
Schematic curves in the lower-right corners of the 3 panels indicate the
temporal evolution of the colours in each plane.  The first and early-second
season photometry shows trends in all three colour-colour planes that are
generally consistent with the star's post-outburst evolution towards later
spectral types and therefore lower temperatures.  The decrease in
temperature is illustrated here by comparison with the locus of blackbody
colours, after adjustment for reddening at the level of $A_{V} = 2.8$
magnitudes found to be appropriate for V838 Mon by Munari, Desidera \&
Henden (2002).  Although blackbody colours are only approximate to the
colours of stellar atmospheres, the early evolution of V838 Mon is seen to
run roughly parallel to the blackbody lines.  The comparison further
highlights the shift towards higher ($K-L$) colours in the middle first
season data noted by Crause et al. (2003), and interpreted by them as
evidence for a short-lived dust formation event.  Thereafter, throughout the
remainder of the second and third seasons, and now with the availability of
early-fourth season photometry, we see the star following complex paths and
possibly the development of loops in the colour-colour diagrams.  We would
hope to elucidate these variations once late-event spectroscopy of the
object becomes more readily available in the literature, e.g. the spectrum
obtained by Evans et al. (2003), announcing the star as an L-type
supergiant, was obtained early in the second season of observations (2002
October 29; AJD 577). Our nearest photometric observation (at AJD 574.6; see
Table 1) records the colours of the star to be $(J-H) = 1.24$, $(H-K) =
0.99$ and $(K-L) = 1.38$.  In Figure 3, this places the star beyond the
position of reddest ($J-H$) colour which likely occurred when V838 was in
conjunction with the Sun in mid-2002, and during a time of decreasing
($J-H$) and ($H-K$) colour.

\begin{figure}
\epsfxsize=8.0cm
\epsffile{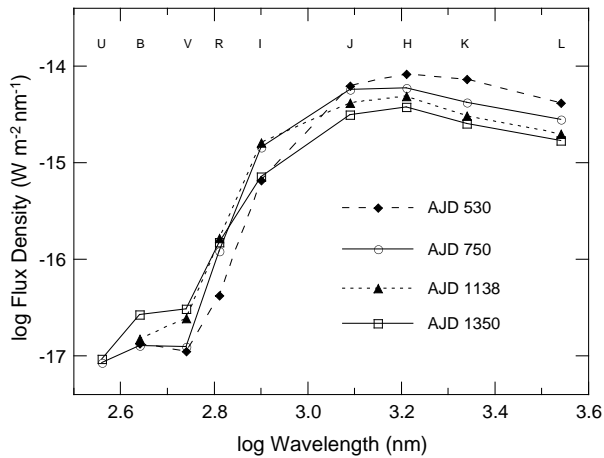}
\caption{The spectral energy distribution of V838 Mon was calculated at 
various epochs.  Since the second observing season, the star has faded in 
the near infrared while brightening in the optical.  The AJD numbers shown 
are average values for the combination of optical and infrared measurements 
used to produce each curve.}
\end{figure}

The brightening seen in the $V$ band during the third and fourth seasons is
in contrast to the near infrared magnitudes that were fading during the same
time period; see Figure 1.  The spectral energy distribution of the star was
calculated at various epochs, corresponding to 2002 September 12 (AJD 530),
2003 April 20 (AJD 750), 2004 May 12 (AJD 1138) and 2004 December 10 (AJD
1350).  These curves are shown in Fig. 4 and illustrate changes in the flux
distribution between the time of the faintest $V$-band measurement (AJD 530)
and our most-recent observations centred at AJD 1350.  In particular note
the $\approx 0.4$ dex (factor of 2.5 in flux, or approximately 1 magnitude)
decrease in flux across the near infrared bands, and the nearly similar
increase in flux seen in the $B$ and $V$ bands.

\section{Light Echo Images}

The SAAO 1.0-m telescope was used to collect {\it BVRI\,} images of the
light echo between 2002 May and 2004 December.  Median averages of 3 or more
frames were produced whenever possible to eliminate cosmic ray strikes and
to increase the signal-to-noise ratio.  When only 1 image was available, the
{\tt IRAF} task {\tt imedit} was used to manually remove cosmic rays and,
when necessary, images were adjusted with {\tt rotate} to compensate for any
differences in the camera orientation from run to run.  Individual images
were finally cropped from 5.3 $\times$ 5.3 arcmin to 2.5 $\times$ 2.5
arcmin.  Fig. 5 shows representative {\it BVRI\,} images chosen to illustrate
the yearly expansion of the light echo.  The 2002--2004 sequences for each
filter were scaled for exposure times and then displayed on the same
intensity scale to facilitate direct comparison.  The {\it BVRI\,} frames
obtained under the best seeing and transparency conditions for each epoch
are available in FITS format and may be downloaded from
ftp://ftp.saao.ac.za/pub/lisa/V838images/.  Dates, exposure times and seeing
details are given in Table 3 and in a `readme' file on the ftp site.

The standard model for the light echo phenomenon involves a short duration
light pulse that propagates outwards and scatters off surrounding dust.  In
the case of V838 Mon however, the star experienced an optical maximum that
lasted about 70 d (see Fig. 1) and during this time the light source also
changed significantly as the star rapidly expanded and cooled.  Since V838
Mon was at its hottest during the 2002 February outburst, the light echo was
brightest in the $U$ and $B$ bands when it was discovered (Henden et al.
2002).  The brightness of the echo decreases due to geometric dilution and
the colour evolution follows that of the star, while also being influenced
by the scattering process that is dependent on the dust properties.

Fig. 5 shows that the echo faded more rapidly at shorter wavelengths than in
the $R$ and $I$ bands and so the dominant factor in the echo colour
evolution appears to be the behaviour of the light source.  For particles
much smaller than the wavelength of light (the Rayleigh regime), the
scattering efficiency is highly wavelength dependent -- proportional to
$\lambda^{-4}$.  If this were the case, we might expect the light echo to
have been much fainter and to have faded faster in the red.  This argument
suggests that the dust particles producing the echo are within the Mie
scattering regime (i.e. the particle sizes are comparable to the wavelength
of light), which is consistent with the properties of interstellar dust grains
(Glass 1999).

\begin{figure*}
\epsfxsize=14.0cm
\epsffile{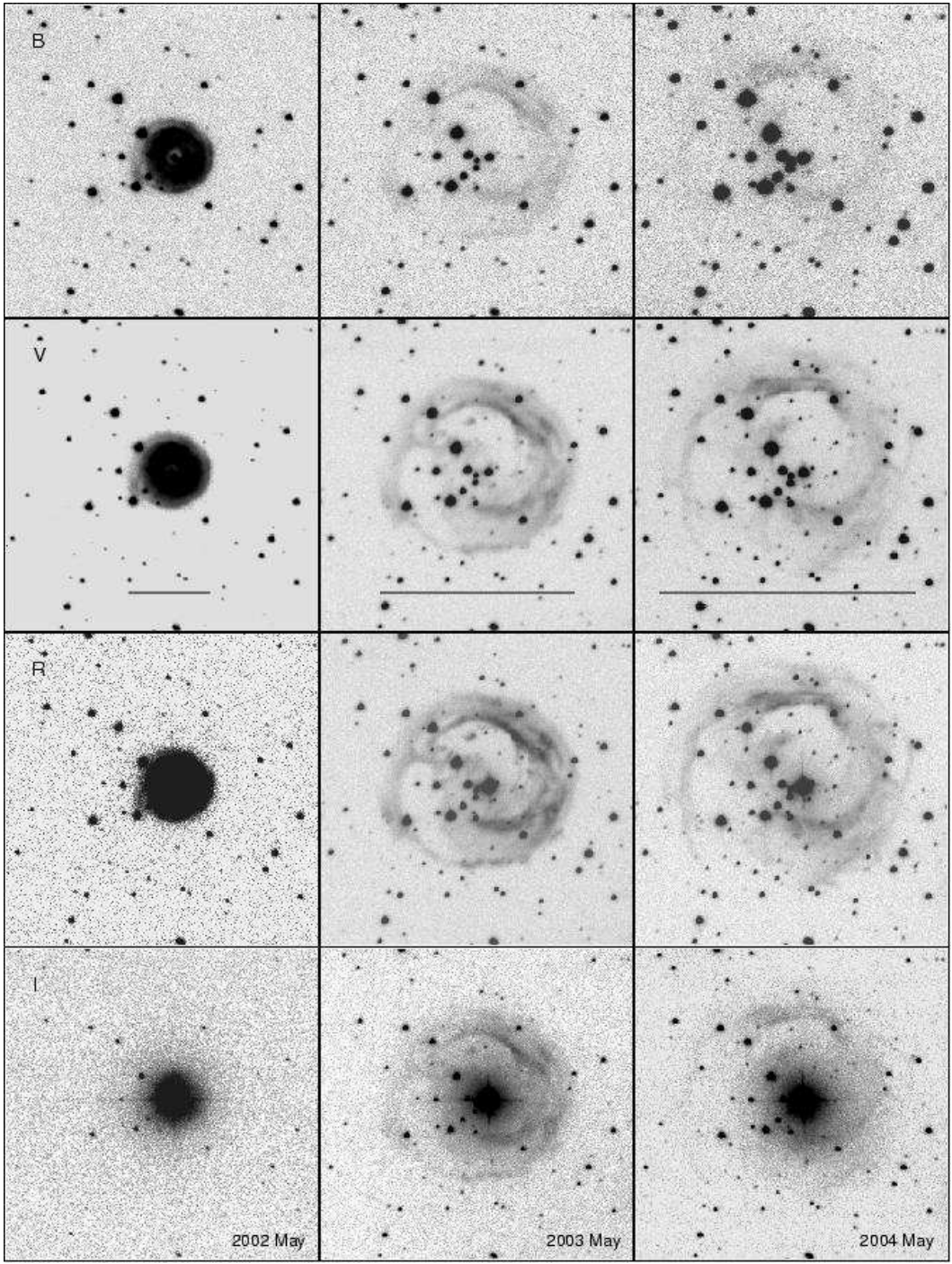}
\caption{$B$, $V$, $R$ and $I$ images of the expanding light echo obtained 
with the SAAO CCD on 2002 May 1 (AJD 396), 2003 May 21 (AJD 781) and 2004 
May 7 (AJD 1133).  Differences in the appearance of stars in the 3 images
for each band are due to seeing variations while the colour evolution of the 
light echo clearly follows that of V838 Mon.  North is up and east to the 
left in each of the 2.5 $\times$ 2.5 arcmin frames.  The measured light echo 
diameter is indicated by a horizontal bar in each of the $V$ images.  See 
Table 3 for further details.}
\end{figure*}

\begin{table*}
\caption{Dates, exposure times (seconds) and seeing measurements (arcsec) for 
the best {\it BVRI\,} images for each epoch.  These were obtained with the SAAO 
1.0-m telescope between 2002 May 1 and 2004 December 11.  During an observing 
run, images for the various filters were often obtained on different nights; 
the UT dates given below are for the $V$ images.  The exact dates for all the 
frames in a given set differ by up to 3 days, details may be found in the 
headers of the files available from ftp://ftp.saao.ac.za/pub/lisa/V838images/.}
\begin{center}
\begin{tabular}{cccccccccc}
\hline
UT Date & AJD & $B$ band & $B$ band & $V$ band & $V$ band & $R$ band & 
$R$ band & $I$ band & $I$ band \\
& & exposure & seeing & exposure & seeing & exposure & seeing & exposure & 
seeing \\
\hline
2002 May 01 & 396 & $1 \times 600$ & 1.4 & $1 \times 300$ & 1.2 & $1 
\times 20$ & 1.3 & $1 \times 2$ & 0.8 \\
2002 Jun 02 & 428 & -- & -- & $1 \times 300$ & 2.1 & $1 \times 100$ & 2.2 & $1 
\times 30$ & 2.0 \\
2002 Sep 16 & 534 & $3 \times 300$ & 2.3 & $3 \times 600$ & 2.0 & $3 
\times 300$ & 1.8 & $3 \times 60$ & 1.7 \\
2002 Oct 31 & 579 & $1 \times 600$ & 2.3 & $1 \times 600$ & 1.9 & $3 
\times 300$ & 1.8 & $5 \times 20$ & 1.4 \\
2003 Jan 26 & 666 & $3 \times 1200$ & 1.5 & $3 \times 900$ & 1.0 & $3
\times 300$ & 1.5 & $5 \times 10$ & 1.1 \\
2003 Apr 19 & 749 & $3 \times 900$ & 1.9 & $3 \times 900$ & 1.8 & $3 
\times 300$ & 1.4 & $9 \times 15$ & 1.5 \\
2003 May 21 & 781 & $3 \times 1200$ & 1.5 & $2 \times 600$ & 1.3 & $3
\times 300$ & 1.1 & $9 \times 10$ & 0.9 \\
2003 Sep 12 & 895 & $3 \times 900$ & 2.3 & $3 \times 600$ & 1.6 & $3
\times 250$ & 1.4 & $5 \times 10$ & 1.0 \\
2003 Nov 03 & 947 & -- & -- & $3 \times 900$ & 1.5 & $3 
\times 300$ & 1.4 & $7 \times 20$ & 1.2 \\
2004 Mar 24 & 1089 & $1 \times 600$ & 1.7 & $1 \times 200$ & 1.5 & $1
\times 200$ & 1.4 & $1 \times 15$ & 1.3 \\
2004 May 07 & 1133 & $1 \times 1800$ & 2.0 & $3 \times 900$ & 1.4 & $3 
\times 300$ & 1.3 & $5 \times 10$ & 0.9 \\
2004 Dec 11 & 1351 & $3 \times 1800$ & 2.0 & $5 \times 1200$ & 1.3 & $3 
\times 600$ & 1.6 & $11 \times 25$ & 1.6 \\
\hline
\end{tabular}
\end{center}
\end{table*}

\section{Estimating the distance from the light echo}

The geometry of a light echo dictates that, provided material remains
present to scatter the light, the echo will appear circular regardless of
the physical distribution of the scattering material; see Tylenda (2004) for
further discussion.  Thus to measure the size of the light echo we fitted a
circle to the outermost edge of the reflection nebula at each epoch and used
the radius as a measure of the echo's angular size.  We measured our $V$
band images as they usually had the highest signal-to-noise ratio in the
nebula.  The data presented in Table 3 show that the $V$ band images had a
mean seeing of $\approx 1.6$ arcsec, with a maximum difference of $\approx
1$ arcsec across the various epochs.  These variations in seeing contribute
to the overall uncertainty of the circular fit, which is dominated by the
inherent difficulty in establishing the `edge' of the light echo.  While the
nebula is sharply defined in the early images, it becomes more diffuse at
later epochs and hence more difficult to measure.  Accordingly, we estimate
the uncertainty in the measured radii to be $\sim 3$ percent, which gives an
absolute error in the radii that scales proportionally to the light echo
radius as it expands with time; see the error bars in Fig. 6.  Table 4 gives
the measured light echo radius at each epoch, along with the day numbers
that denote the interval between each observation and the time of the star's
maximum brightness (2002 February 5; AJD 311), when the light echo is
believed to have been triggered.

In measuring the radii, we found that the centre of the light echo
systematically shifted away from V838 Mon, towards the NE (in the upper-left
direction in Fig. 5).  This shift tends to make the echo seem asymmetric,
but merely off-setting the circle being fitted compensates for this effect. 
Such a shift suggests that the scattering dust is in the form of a sheet
that is inclined relative to the line-of-sight; the measured Right Ascension
and Declination offsets in units of arcsec are listed in Table 4.  These
offset values have uncertainties of a few arcsec, but clearly follow a
linear trend that allows us to derive the inclination angle of the implied
sheet once an estimate of the distance to the star has been obtained (for
details see Tylenda 2004).

Tylenda (2004) modeled the light echo evolution for dust distributed in thin
interstellar sheets and in thin circumstellar shells.  Both cases initially
produce an expanding echo that cannot be distinguished by model fits to the
observations.  However, the models predict a divergence with time -- shells
of circumstellar dust will eventually result in the light echo contracting
and disappearing, while expansion due to interstellar dust will decelerate
but still continue as the echo gradually fades.  Resolving the nature of the
scattering material justifies the continued interest in imaging of the
nebula.  For both the dust distribution models considered here, the angular
expansion of the light echo can, in principle, be used to determine the
stellar distance and the distance of the scattering material from the star.

Fig. 6 shows our measured radii versus the time since the peak of the
2002 February outburst (AJD--311).  To these data we fitted models for
scattering by a single thin sheet ($+$ sign) and a single thin shell ($-$
sign) of material, making use of the equation $\theta = 1/d$ $\sqrt{2rct \pm
c^2t^2}$, where $\theta$ is the angular radius of the echo, $d$ is the
distance to the star, $r$ is the distance between the star and the dust, $c$
is the speed of light and $t$ is the time since the peak of the outburst. 
This form is equivalent to that given by van Loon et al. (2004), other than
invoking the small angle approximation as the angles involved will always be
sufficiently small that tan $\theta \approx \theta$.

$\chi^2$ minimisation yielded $r$ and $d$ values of $4.8 \pm 1.8$ pc
and $8.9 \pm 1.6$ kpc for the sheet model and $9.4 \pm 9.8$ pc and
$12.0 \pm 6.5$ kpc for the shell model, respectively.  The model-derived
results are sensitive to uncertainties, or errors, in the measured
radii\footnote{van Loon et al. (2004) measured the diameter of the light
echo in images obtained with various telescopes, including the {\it HST} and
the SAAO 1.0-m, and attempted to fit thin sheet and thin shell models to
these data.  As they were unable to find solutions for either model that
fitted both the early and later epochs, they relaxed the assumption of a
thin scattering medium and derived a lower limit of $> 5.5$ kpc for the
distance to V838 Mon.  However, comparison between the diameters given by
van Loon et al. (see their table 1) and the radii listed by Tylenda (2004;
see table 1 and 2) and in Table 4 of this paper reveals that the van Loon et
al. diameters are in error, underestimating the actual echo diameters by a
factor of $\sim 2.5$.  We also note that the UT dates estimated for their
third and sixth epochs (SAAO images) are in error by 16 and 1 d
respectively.}, with the shell case being additionally complicated by the
need to provide a reasonable initial value for the radius $r$ so as to avoid
obtaining a negative term within the square root term of the expansion
equation.

Tylenda (2004) discusses the intrinsic uncertainty of the fits due to the
shallow, extended nature of the $\chi^2$ minimum, but it is encouraging to
note that both models produced values for the distance to V838 Mon that are
consistent with lower limits proposed by various authors (e.g. Kipper et al.
2004, Wisnewski et al. 2003, Bond et al. 2003) and in agreement with values
for the distance of 9.2 kpc and $8 \pm 2$ kpc derived recently by Lynch et
al. (2004) and Tylenda (2004), respectively. Fig. 6 suggests that the
expansion of the echo may be entering a regime where it has become possible
to distinguish between the two dust distribution models owing to the
significant divergence in the model-predicted light echo radii.

Having obtained a distance $d = 8.9 \pm 1.6$ kpc from the sheet model, we
estimated the inclination angle ($\alpha$) of the dust sheet's normal to the
line-of-sight.  Following Tylenda (2004), we applied a linear fit to the
measured RA-offset of the echo-centre versus time (since outburst) and
derived a slope of $\dot{\theta}_c = 1.01\times10^{-2}$ arcsec\,day$^{-1}$. 
Since tan $\alpha = \dot{\theta}_{c}d/ct$, appropriate unit conversions
(arcsec\,day$^{-1}$ into radians\,day$^{-1}$ and then radians into degrees)
yielded $\alpha \approx 27^{\rm o}$.

\begin{table}
\caption{Measured light echo radii for each of the $V$ band images
and their respective dates, as well as a time representing the
number of days since the peak of the 2002 February outburst (AJD--311).
Right Ascension ($\alpha$) and Declination ($\delta$) offsets between 
V838 Mon and the centre of the echo are also given.}
\begin{center}
\begin{tabular}{cccccc}
\hline
UT Date & AJD & Time & Radius & $\alpha$-offset & $\delta$-offset \\
        & & (d) & (arcsec) & (arcsec) & (arcsec) \\
\hline
2002 May 01 & 396 & 85  & 19.5 & $-1$ & 1 \\
2002 Jun 02 & 428 & 117 & 21.4 & $-0$ & 0 \\
2002 Sep 16 & 534 & 223 & 31.9 & $-2$ & 0 \\
2002 Oct 31 & 579 & 268 & 34.1 & $-2$ & 2 \\
2003 Jan 26 & 666 & 355 & 39.4 & $-3$ & 2 \\
2003 Apr 19 & 749 & 438 & 44.3 & $-4$ & 3 \\
2003 May 21 & 781 & 470 & 46.5 & $-5$ & 3 \\
2003 Sep 12 & 895 & 584 & 51.8 & $-6$ & 3 \\
2003 Nov 03 & 947 & 636 & 53.9 & $-7$ & 3 \\
2004 Mar 24 & 1089 & 778 & 58.6 & $-7$ & 5 \\
2004 May 07 & 1133 & 822 & 61.4 & $-8$ & 7 \\
2004 Dec 11 & 1351 & 1040 & 70.1 & $-10$ & 9 \\ 
\hline
\end{tabular}
\end{center}
\end{table}

\begin{figure} 
\epsfxsize=8.0cm 
\epsffile{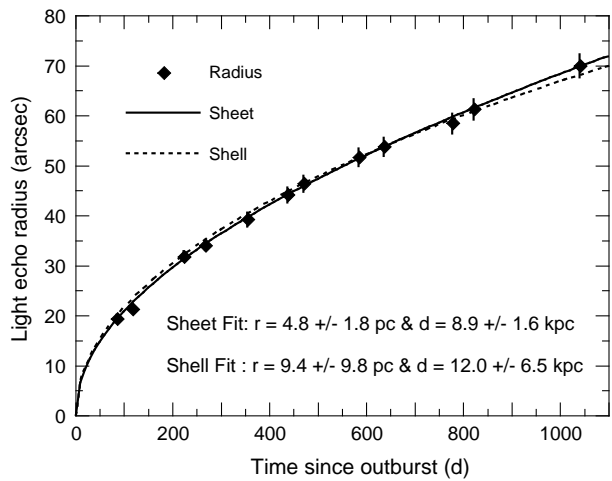} 
\caption{Measured light echo radius (arcsec) versus the time (d) since the peak
outburst in 2002 February.  Best-fit models for scattering material distributed
in a thin sheet and a thin shell are shown and the values for the stellar 
distance, $d$, and star-dust separation, $r$, derived from each fit are listed. 
Error bars indicate the estimated 3 percent uncertainty in the measured echo 
radii.}
\end{figure}

\section{Discussion}

Lynch et al. (2004) model $0.8-2.5$, $2.1-4.6$ and $3-14 \mu$m infrared
spectra with a central star and a 2-component expanding circumstellar shell. 
They treat H$_{2}$O as the dominant molecular absorber and emitter in the
envelope which attenuates stellar emission in the near infrared and emits
thermally in the infrared.  H$_{2}$O and CO transmission through the shell
dominates in the $2.5-4.5 \mu$m region and the photosphere makes a minimal
contribution beyond $4.5 \mu$m.

Their model does not include silicate dust and they point out that the 2002
February/March extinction can be explained with a combination of molecular
Rayleigh scattering and molecule formation.  However, as early as 2002
January, their spectra showed a slightly elevated $10 \mu$m region and by
2003 January they report that the $8-13 \mu$m flux had increased 10 times
more than that in the near infrared.  While the development of this infrared
excess could be due to gas-phase molecules in the circumstellar envelope, a
large amount of dust-precursor material was present in the form of molecular
SiO and in 2003 February a fairly strong feature near $10 \mu$m was
tentatively identified as silicate emission with a central absorption
feature (Sitko et al. 2003).  The timing of this observation roughly
coincides with the end of the broad peak noted in the infrared light curves
(Fig. 1) and the time of the maximum ($K-L$) colour of V838 Mon (Fig. 2c). 
This confirms that at least some of the infrared photometric activity was
caused by the production of dust, although the extent to which dust
formation occurred throughout the outburst remains in question.

Light echoes have been detected in several extra-galactic supernovae
(Chevalier 1986), the best studied being that of SN 1987A (Xu, Crotts \&
Kunkel 1995).  However, prior to V838 Mon, Nova Persei 1901 (Couderc 1939)
was the only Galactic variable known to have developed a light echo.  

While most observed light echoes have been due to interstellar dust, it is
conceivable that the scattering could take place within a circumstellar
envelope.  The evolutionary status of V838 Mon remains uncertain, but
circumstellar dust would be expected if the star has undergone multiple
outbursts in the past.  The discovery of extended cool shells surrounding
V838 Mon reported by van Loon et al. (2004) and detections by both 2MASS and
IRAS (Kato 2002) may imply that the progenitor was an evolved object.

However, if V838 Mon is indeed kinematically associated with a B3V companion
(Desidera \& Munari 2002, Wagner \& Starrfield 2002), the system has to be
young, indicating a massive progenitor for the outbursting component.  This
outcome is consistent with Tylenda et al. (2004) who claim the progenitor
was most likely a $5-10$ M$_{\odot}$ dwarf, and who also reject the link
between V838 Mon and the extended cool emission found by van Loon et al.
(2004).  Reddening of the B3V star indicates that the system is
$\sim 10.5$ kpc distant (Munari et al. 2002) and although the star is
situated towards the Galactic anti-centre, its Galactic latitude of
$1.05^{\rm o}$ places it within 200 pc of the plane, consistent with the
observed extinction being due to interstellar material.

Also, Wisnewski et al. (2003) and Desidera et al. (2004) conclude that the
polarization observed in V838 Mon is mostly of interstellar origin.  A small
intrinsic component was present during the 2002 February maximum, but
decreased rapidly after that.  Furthermore, deep polarimetric imaging of the
light echo (Desidera et al. 2004) shows a dipole structure that would not be
consistent with scattering from a homogeneous circumstellar medium.

While the real dust distribution that results in the light echo is likely to
be complex, Fig. 6 shows that the observed light echo expansion can be
reproduced by simple models that assume a thin sheet or a thin shell of
scattering material.  Tylenda (2004) analysed the {\it HST} images of the
light echo and found that a sheet of material matched the measured radii
better than a shell model could.  Only a shell of implausibly large radius,
hence approximating a flat sheet of material, could be made to fit the data. 
Our larger fit-errors for the shell (105 and 54 percent for $r$ and $d$
respectively, compared to 37 and 18 percent for the sheet) also indicate
that the sheet model more successfully represents the data.

We expect the nature of the scattering material will be established by the
end of the fourth observing season as the sheet and shell models will have
separated significantly by that time and hence allow discrimination
between the 2 possibilities.  With the echo fading due to geometric
dilution, high signal-to-noise images will be required to avoid
under-estimating the radius as that would bias fits towards the shell model.

Future measurements of the radius will also help to refine the distance as
later points will improve the model fit and hence provide better estimates
of $d$ and $r$.  At a much later stage, the observed shift between the echo
centre and the star could be used, in conjunction with the radius
measurements, as an alternative method to determine the star's distance
(Tylenda 2004).

\section{Summary}

V838 Mon seemed to have settled down after the dramatic 2002 events, but the
photometry presented in Tables 1 and 2 indicate that the properties of the
object continued to evolve (see Fig. 1).  During the second season, the star
remained faint in the $V$ band but brightened by $\sim 0.8$ mag in $R$ and
$I$ and developed a broad, flat bump in the $JHKL$ bands that peaked around
AJD 600.  The $JKL$ infrared magnitudes then faded slowly during the third
season while the $H$ and $I$ bands remained fairly constant and the $V$ band
began brightening.  The complex infrared evolution shown in Figs 2 and 3 and
changes in the overall flux distribution shown in Fig. 4 appear consistent
with a dust dispersal scenario.

Fig. 5 consists of sample {\it BVRI\,} light echo images spanning a 2 year
interval; see Table 3 for all image details.  Our current set of images may
be downloaded from the SAAO ftp site; see Section 4 for details.

We measured the radius of the expanding light echo in our $V$ band images
and fitted models for a thin sheet and a thin shell of scattering material. 
While the thin sheet model gave a somewhat better fit to the light echo
radii, both models yield distances ($8.9 \pm 1.6$ kpc and $12.0 \pm 6.5$ kpc
for the sheet and shell respectively) that are consistent with estimates
from other methods -- see Fig. 6 and Table 4.  However, the colour evolution
of the light echo, the higher quality of the model fit for a sheet of dust,
the observed shift in the echo centre relative to the star, results from
polarimetric studies and the position of the variable only slightly above
the Galactic plane all suggest that the dust illuminated by the outburst of
V838 Mon is more likely to be of interstellar origin.

Establishing the nature of the dust may provide evolutionary information
while the distance is essential for accurately determining the absolute
luminosity of this mysterious variable.  Continued imaging thus remains a
priority; with time the dust distribution will be revealed and the better
constrained fit will improve the distance estimate.

\section*{Acknowledgments}

Our thanks to the SAAO Time Allocation Committee for generous 1.0-m time
allocations, to Romuald Tylenda for helpful discussions about the light echo
and to Jacco van Loon for his constructive referee's report that resulted in
significant revision and improvement of this paper.  WAL acknowledges
research support from UNSW@ADFA Faculty Research Grants and Special Research
Grants.

\end{document}